\pdfoutput=1

\documentclass[pra,twocolumn,showpacs]{revtex4-1}
\usepackage{amsmath}
\usepackage{amsfonts}
\usepackage{amssymb}
\usepackage{graphicx}

\begin{document}

\title{Intensity and amplitude correlations in the fluorescence from atoms with interacting Rydberg states}
\author{Qing Xu and Klaus M{\o }lmer}
\email{moelmer@phys.au.dk}
\affiliation{Department of Physics and Astronomy, Aarhus University, Ny
Munkegade 120, DK-8000 Aarhus C, Denmark.}
\date{\today}

\begin{abstract}
We explore the fluorescence signals from a pair of atoms driven towards Rydberg states on a three-level ladder transition. The dipole--dipole interactions between Rydberg excited atoms significantly distort the dark state and electromagnetically induced transparency behavior observed with independent atoms and, thus, their steady state light emission. We calculate and analyze the temporal correlations between intensities and amplitudes of the signals emitted by the atoms and explain their origin in the atomic Rydberg state interactions.
\end{abstract}

\maketitle

\renewcommand{\c}[0]{\ensuremath{\hat{c}}} \renewcommand{\H}[0]{\ensuremath{%
\hat{H}}} \renewcommand{\Im}[0]{\ensuremath{\mathrm{Im}}} %
\renewcommand{\L}[0]{\ensuremath{\hat{L}}} \renewcommand{\Re}[0]{%
\ensuremath{\mathrm{Re}}} \renewcommand{\vec}[1]{\ensuremath{\mathbf{#1}}}


\section{Introduction}
The measurement of temporal field correlation functions plays a
fundamental role in the demonstration of nonclassical properties of light and in the analysis of the underlying dynamics of quantum emitters \cite{Glauber}. Field--field correlations and
intensity--intensity correlations, reflecting wave and
particle properties of light, respectively, have thus been studied for the emission by a single atom \cite{Mollow,
Kimble, Huang, Swain} and by few atoms \cite{Ficek}, while particle and wave aspects have been jointly addressed for amplitude--intensity correlations in the emission by a single atom \cite{Foster, Denisov, Marquina-Cruz, Gerber}.
While the correlation functions can be expressed as two-time averages in the Heisenberg picture and can be calculated by use of the
quantum regression theorem \cite{Gardiner, Carmichael}, a simple intuition for their behavior can be obtained by alternatively considering the conditional dynamics of the light emitting system. \textit{E.g.}, after a photon counting event, the atom is put in its ground state, and its subsequent transient dynamics shows damped population and coherence oscillations, which are naturally reflected by the temporal modulation of the light emission characteristics.

In this article, we consider the intensity and amplitude correlations in the fluorescence emitted by a pair of atoms that are both excited on three-level ladder transitions under electromagnetically induced transparency  (EIT) conditions \cite{Harris}. Such atoms will emit only very little light, and since an emission event on the short-lived lower transition is accompanied by a quantum jump of the atom into the ground state, both the field intensity and amplitude will show strong transient dynamics around the rare counting events \cite{Marquina-Cruz,Xu}. In the absence of interactions a pair of atoms will emit light in a mutually uncorrelated manner, but as shown in \cite{Pritchard1}, a strong interaction between two atoms which are both in the upper exited state, leads to correlations between the emitted intensities.

The physical situation can be implemented with the use of highly excited Rydberg states as the upper state in the level scheme shown in Fig. 1. Such states have long lifetimes, and due to their large dipole moments, two nearby atoms experience a large interaction energy shift, which in turn shifts the resonance condition for the excitation of both atoms. This effective detuning may be large enough to prevent the excitation (the Rydberg blockade \cite{Jaksch,Urban, Gaetan}), or it may merely detune the state with two excited atoms. In either case, the EIT properties are distorted and the atoms emit more light \cite{Pritchard1, Ates, Petrosyan, Petrosyan2, Pritchard}.

Rydberg atoms have received much recent attention due to their potential application in quantum information processing \cite{Saffman} where the Rydberg blockade mechanism can be used to mediate controlled quantum gates \cite{Jaksch, Lukin1}.  EIT arises from the destructive interference between different absorption and emission processes \cite{Harris}, beneficial for
slowing of light \cite{Hau} and quantum nondemolition interactions \cite{Imoto}, while the Rydberg blockade in
combination with EIT is attracting interest as it paves the road for strong cooperative optical nonlinearities \cite{Ates, Petrosyan,Pritchard, Stanojevic, Dudin, Peyronel, Hofmann}.

We shall characterize the field correlations from a pair of  atoms, exploring how they may on the one hand serve as a  probe of the Rydberg interactions, and, on the other hand as a potential heralded source of nonclassical radiation. In Sec. II, we introduce the master equation of our interacting atoms and we calculate and analyze two-time correlation functions of the emitted radiation. In Sec. III, we turn our attention to \emph{three-time} correlation functions, addressing in particular how the field amplitude emitted by one atom behaves between counting events and quantum jumps of the two emitters. In Sec. IV, we conclude and summarize our results.

\section{Density matrix and two-time correlations}
In this section we analyze the two-time correlations using the master equation and quantum regression theorem. In
Sec. \ref{IIA} we present the master equation of two three-level ladder atoms subject to Rydberg--Rydberg interactions. The intensity--intensity correlations and intensity--amplitude correlations are discussed in Sec. \ref{IIB} and in Sec. \ref{IIC}, respectively.
\subsection{The physical system and the master equation}
\label{IIA}
\begin{figure}[tbp]
\centering%
\includegraphics[bb=171 510 432
749,width=0.65\columnwidth,keepaspectratio]{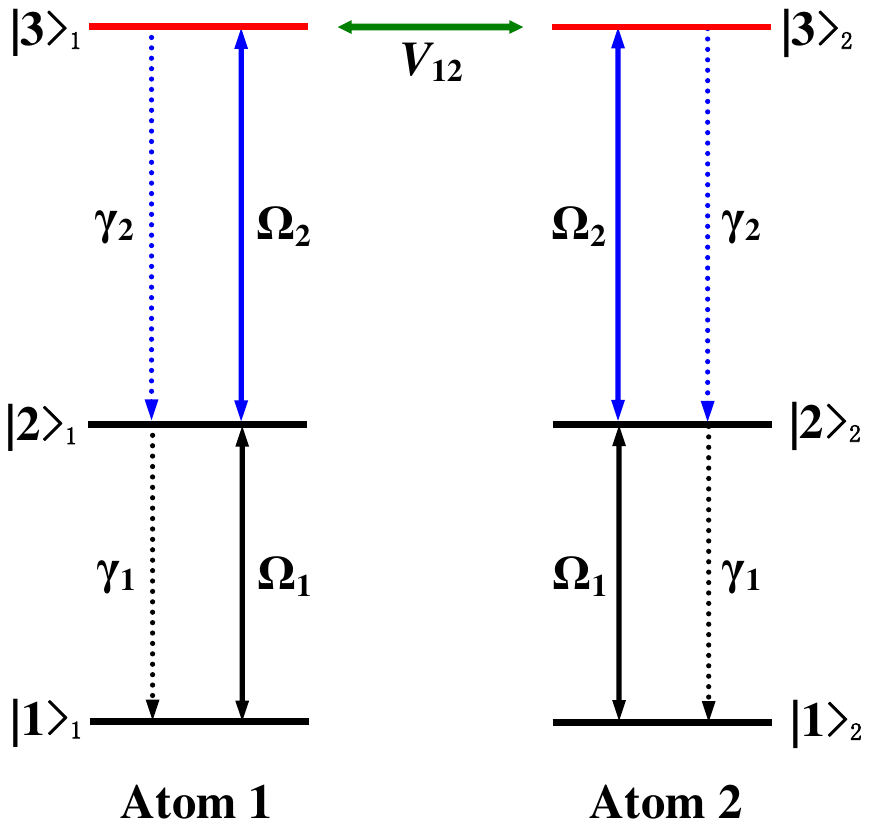}
\caption{(Color online) Level diagram of two three-level ladder atoms with a lower state $|1\rangle $, a
short-lived intermediate state $|2\rangle $, and a long-lived Rydberg state $|3\rangle $.
The atoms are coupled to each other by the dipole--dipole interaction $V_{12}$ between the
Rydberg states. In both atoms the lower transition, with a line width $\gamma_1$ due to atomic decay, is driven by a resonant probe field with Rabi frequency $\Omega _{1}$, and the upper transition subject to decay $\gamma_2$ and dephasing $\gamma_{\mathrm{ph}}$ of the upper level, is coupled to a resonant laser field with Rabi frequency $\Omega _{2}$.}
\label{ladder}
\end{figure}
Our physical system involves two
atoms excited by a pair of laser fields towards a Rydberg state in the ladder configuration, shown schematically in Fig. \ref{ladder}.
The time evolution of the atomic system can be described by the reduced atomic density matrix $\rho$, which in the Schr\"{o}dinger picture obeys a linear master equation,
\begin{equation}
\dot{\rho}=\mathcal{L}\rho,  \label{ME}
\end{equation}%
where \begin{equation}
\mathcal{L}\rho=\frac{1}{\mathrm{i}\hbar }[H,\rho ]+\sum_{j,k}C_{k}^{(j)}\rho
C_{k}^{(j)\dagger }-\frac{1}{2}\left\{ C_{k}^{(j)\dagger }C_{k}^{(j)},\rho
\right\} .  \label{lindblad}
\end{equation}%
$H$ is the total Hamiltonian of the system and takes the form%
\begin{equation}
H=H_{1}\otimes I_{2}+I_{1}\otimes H_{2}+H_{\mathrm{dd}},
\end{equation}%
with single-atom Hamiltonians $H_{j}$ ($j=1,2$)
\begin{eqnarray}
H_{j} &=&-\frac{\hbar }{2}(\Omega _{1}\sigma _{21}^{(j)}+\Omega _{2}\sigma
_{32}^{(j)}+\text{\textrm{H.c.}}),
\end{eqnarray}%
and the dipole--dipole interaction $H_{\mathrm{dd}}$  between the two
Rydberg states
\begin{equation}
H_{\mathrm{dd}}= V_{12}\sigma _{33}^{(1)}\sigma _{33}^{(2)}.
\end{equation}%
Here $\sigma _{kl}^{(j)}=|k\rangle _{jj}\langle l|$ are the atomic
operators ($j=1,2$; $k,l=1-3$), of the $j$th atom, $\Omega _{1,2}$ are the
laser Rabi frequencies, and $V_{12}$ is the strength of the dipole--dipole
interaction. The quantum jump operators $%
C_{k}^{(j)}$ of the $j$th atom account for dissipative couplings to the environment of the system, and take the form $C_{1}^{(j)}=\sqrt{\gamma _{1}}\sigma
_{12}^{(j)}$ and $C_{2}^{(j)}=\sqrt{\gamma _{2}}\sigma _{23}^{(j)}$ with
the respective decay rates $\gamma _{1,2}$, and $C_{3}^{(j)}=\sqrt{\gamma _{\mathrm{ph}}}%
(\sigma _{33}^{(j)}-\sigma _{22}^{(j)}-\sigma _{11}^{(j)})$ with a dephasing
rate $\gamma _{\mathrm{ph}}$ of the high-lying Rydberg state with respect to
the two lower states.

We recall that in the absence of decay and dephasing of the upper level $|3\rangle$, a
resonantly driven three-level ladder atom has a dark eigenstate, $|\mathrm{D}\rangle =\frac{1}{\Omega _{\mathrm{R%
}}}(\Omega _{1}^{\ast }|3\rangle -\Omega _{2}|1\rangle )$, where
$\Omega _{\mathrm{R}}=\sqrt{|\Omega _{1}|^{2}+|\Omega _{2}|^{2}}$. The system evolves into this state, which has no component of the intermediate, short-lived state, and hence emits no photons. For two such atoms close enough to each other, the Rydberg blockade provides a detuning of the transitions that deteriorates the dark state of the atoms. Rather than populating a product state of the dark atomic states, $|\mathrm{DD}\rangle$, an Atom 1 Rydberg state component restricts Atom 2 to explore the resonant two-level transition $|1\rangle \leftrightarrow |2\rangle$ and emit fluorescence photons, and \textit{vice versa} \cite{Petrosyan2}. Fluorescence from a single atom is known to show antibunching, but the signal from the atom pair, instead, shows significant bunching \cite{Pritchard1}. This is qualitatively explained by the system exploring the eigenstates of the full Hamiltonian $H$ (3), which have only single occupancy of the Rydberg state $|3\rangle$, but which have also acquired population of the product state component $|22\rangle$ \cite{Moller}. When a photon is emitted by one of the atoms, it heralds an increased probability for the other atom to be in the short-lived excited state and emit a subsequent photon.

\subsection{Intensity--intensity correlation functions}
\label{IIB}
The correlations in photon counting from atoms $i$ and $j$ are evaluated as normal ordered correlation functions of the atomic dipole lowering and raising operators \cite{Glauber,Carmichael}, and are conveniently normalized by the product of the mean emission rates to yield the correlation function
\begin{equation}
g_{ij}^{(2)}(\tau )=\lim_{t\rightarrow \infty }\frac{\langle \sigma
_{21}^{i}(t)\sigma _{22}^{j}(t+\tau )\sigma _{12}^{i}(t)\rangle }{\langle
\sigma _{22}^{i}(t)\rangle \langle \sigma _{22}^{j}(t+\tau )\rangle },
\end{equation}%
with $i,j=1,2$.
\begin{figure}[tbp]
\centering%
\includegraphics[bb=72 321 503
678,width=0.95\columnwidth,keepaspectratio]{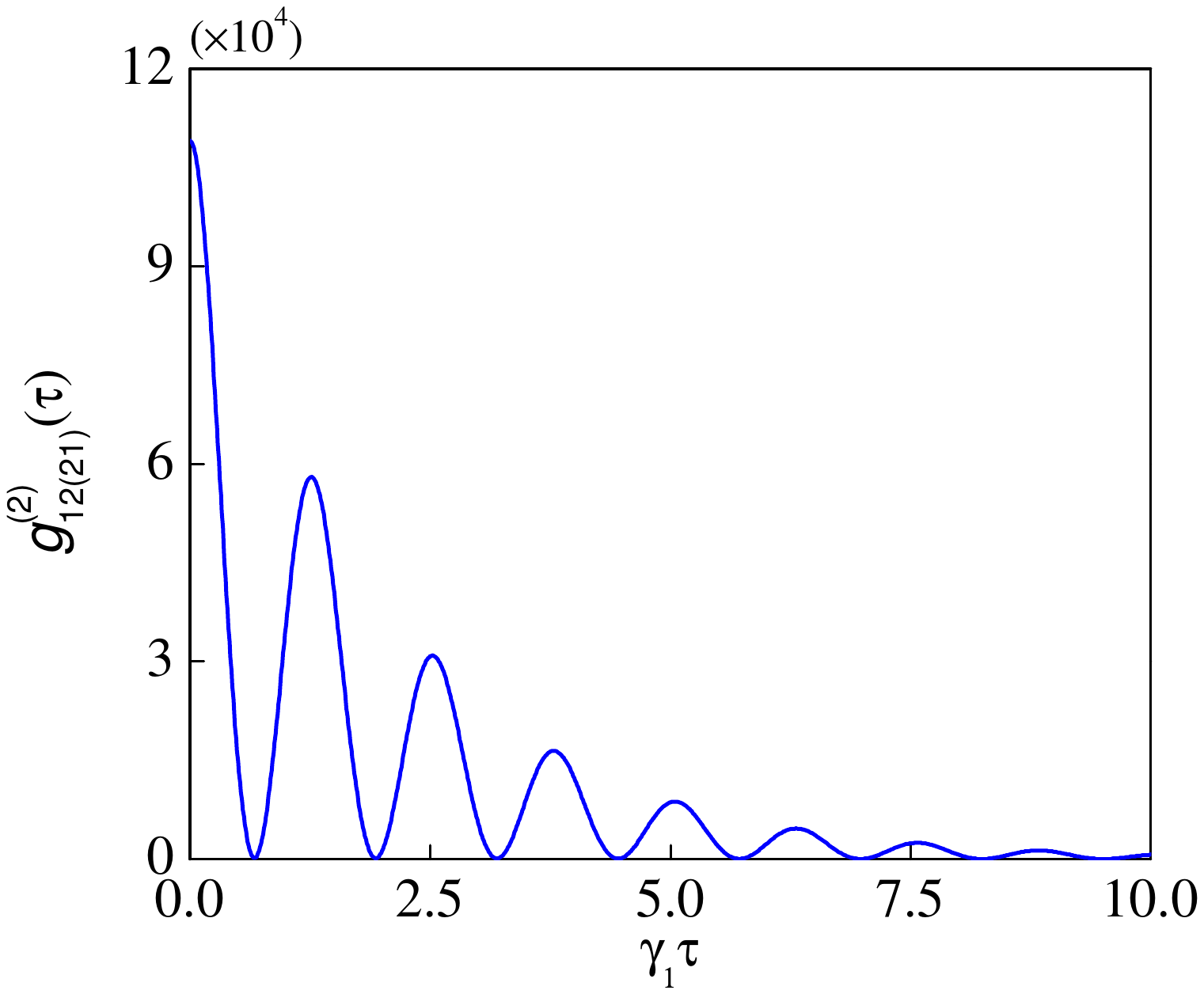}
\caption{(Color online) Intensity--intensity cross correlation on the lower
transitions of two interacting Rydberg atoms. The parameters are: $\Omega
_{1}=0.2\protect\gamma _{1}$, $\Omega _{2}=5\protect\gamma _{1}$, $V_{12}=%
\hbar\protect\gamma _{1}$, $\protect\gamma _{2}=1\times 10^{-4}%
\protect\gamma _{1}$, and $\protect\gamma _{\mathrm{ph}}=1\times 10^{-4}%
\protect\gamma _{1}$.}
\label{II}
\end{figure}
This correlation function is readily calculated by the master equation and the quantum regression theorem \cite{Gardiner,Carmichael}, and it is illustrated for a characteristic example in Fig. \ref{II}, see also \cite{Pritchard1}.
We observe that the cross correlation is subject to damped Rabi oscillations with a period of
$2\pi /\Omega _{\mathrm{R}}$. This is because the quantum regression theorem provides a linear set of equations for all two-time correlators with exactly the same coefficients \cite{Gardiner, Carmichael} as the master equation (1), but we can also interpret the correlation function in (6) as a consequence of the measurement back action. Conditioned on the detection event the joint state of the two atoms undergoes a quantum jump by the operator $\sigma_{12}^i$ at time $t$, and the subsequent $t+\tau$ dependent emission rate from atom $j$ is given by its transient excited state population.  The formal equivalence of Glauber's photodetection theory and the quantum theory of measurements is further elaborated in Ref. \cite{Xu}.

\subsection{Intensity--amplitude correlation functions}
\label{IIC}
\begin{figure}[tbp]
\centering%
\includegraphics[bb=95 43 503
772,width=0.9\columnwidth,keepaspectratio]{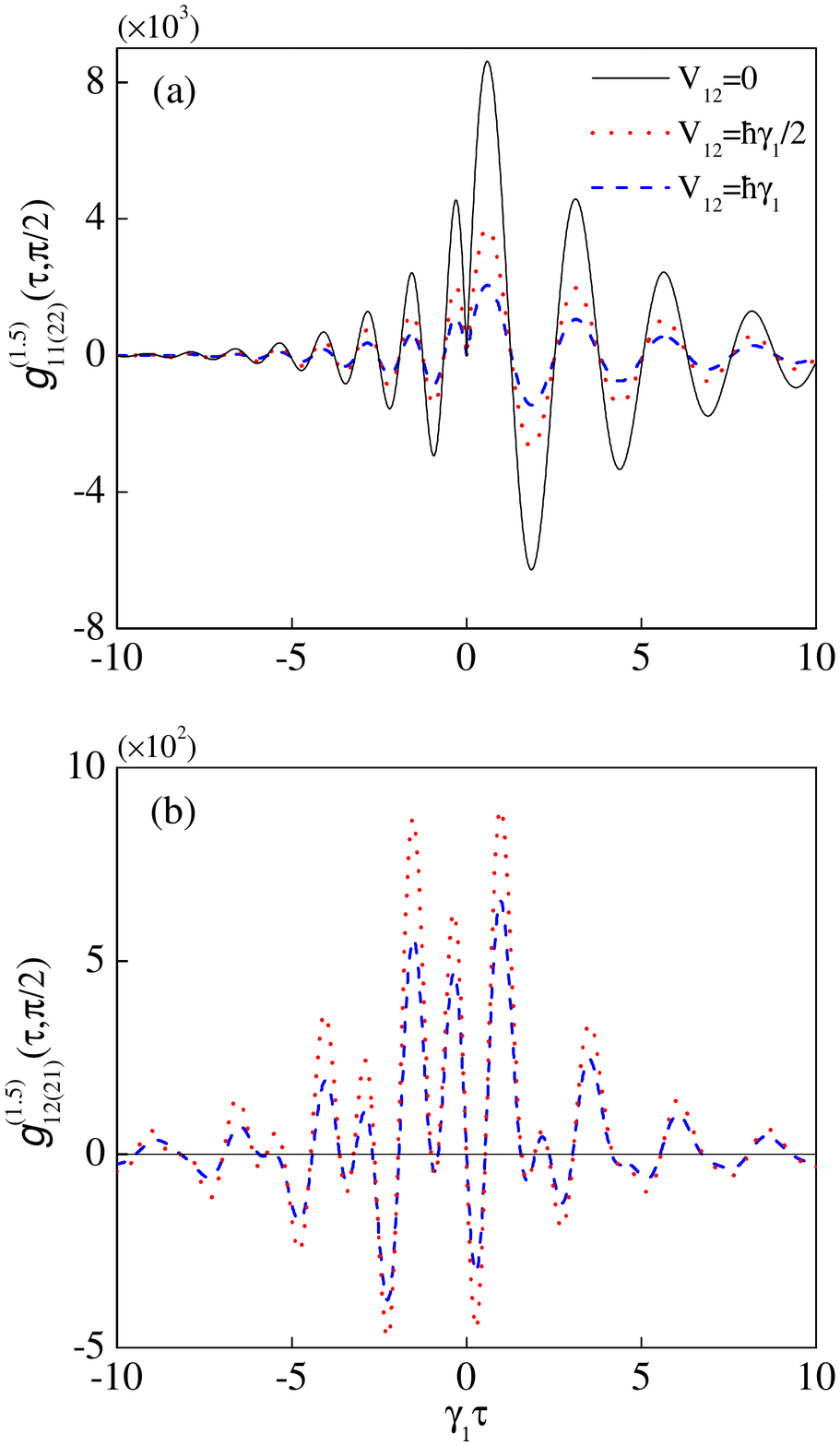}
\caption{(Color online) Amplitude--intensity autocorrelation $%
g_{11(22)}^{(1.5)}(\protect\tau ,\protect\pi /2)$ (a) and cross correlation $%
g_{12(21)}^{(1.5)}(\protect\tau ,\protect\pi /2)$ (b) on the lower
transitions of two three-level atoms with excited state (Rydberg) interaction $V_{12}=0$ (black solid line), $V_{12}=%
\hbar\protect\gamma _{1}/2$ (red dotted line) and $\hbar\protect\gamma _{1}$ (blue
dashed line). The other parameters are the same as in Fig. \protect\ref%
{II}.}
\label{AI}
\end{figure}

In \cite{Foster, Denisov, Marquina-Cruz, Gerber}, it was demonstrated that combined photon counting and homodyne detection of the light emitted by an atom, also shows mutual temporal correlations. Phase sensitive homodyne detection may have near unit efficiency, it holds the potential to reveal more information than photon counting about the emitter dynamics, and, through tomography it may be used to fully reconstruct general quantum states of light \cite{Tomography}. Ensembles of atoms with Rydberg interactions exhibit interesting quantum dynamics and may yield sources of nonclassical radiation, and in this work, we will thus address the influence of the Rydberg interactions on the intensity--amplitude autocorrelation and cross correlations.

The detection of the field quadrature variable with phase $\theta_j$ from atom $j$ at time $t+\tau$ is correlated with the counting of a photon from atom $i$ at the earlier time $t$ as quantified by the steady state ($t\rightarrow \infty$) correlation function $(\tau > 0)$:
\begin{equation}
g_{ij}^{(1.5)}(\tau ,\theta _{j})=\lim_{t\rightarrow \infty }\frac{\mathrm{%
Re\,}[\langle \sigma _{21}^{i}(t)\sigma _{21}^{j}(t+\tau )\,\mathrm{e}^{\mathrm{i}\theta _{j}}\sigma
_{12}^{i}(t)\rangle]}{\langle \sigma
_{22}^{i}(t)\rangle \,\mathrm{Re\,}[\langle \sigma _{21}^{j}(t+\tau )\rangle
\,\mathrm{e}^{\mathrm{i}\theta _{j}}]}.
\end{equation}%
We shall also determine the correlations for negative $\tau$, \textit{i.e.}, for field amplitude measurements preceding the count events, and due to the normal ordering requirement \cite{Glauber}, the correlation function is here given by a different expression ($\tau < 0$):
\begin{equation}
g_{ij}^{(1.5)}(\tau ,\theta _{j})=\lim_{t\rightarrow \infty }\frac{\mathrm{%
Re\,}[\langle \sigma _{21}^{j}(t)\,\mathrm{e}^{\mathrm{i}%
\theta _{j}}\sigma _{22}^{i}(t-\tau )\rangle]}{\langle \sigma _{22}^{i}(t-\tau )\rangle \,%
\mathrm{Re\,}[\langle \sigma _{21}^{j}(t)\rangle\,\mathrm{e%
}^{\mathrm{i}\theta _{j}}]}.
\end{equation}%

In Fig. \ref{AI}(a), we show the intensity--amplitude correlation function in Eqs. (7) and (8) for one atom, i.e., $i=j$, in the presence of the other atom. The correlation function is asymmetric with faster oscillations for negative than for positive times $\tau$. The interpretation of this difference was developed for the emission by a single atom in \cite{Xu}: After a photon detection event, the atom recommences its Rabi oscillation from the ground state $|1\rangle$. This is a linear combination of the dark state $|\mathrm{D}\rangle$ and two \textquotedblleft bright\textquotedblright ~eigenstates states separated in energy from $|\mathrm{D}\rangle$ by $\pm \hbar\Omega_{\mathrm{R}}/2$, and hence  physical expectation values oscillate naturally at the frequency $\Omega_{\mathrm{R}}/2$. By contrast, to be consistent with the emission of a photon at time $t$, the evolution prior to the photon count event correlates with the excited state $|2\rangle$ which has a vanishing overlap with the dark state and is thus composed only of the bright states separated by $\hbar\Omega_{\mathrm{R}}$. This causes the atomic observables and hence the field amplitude to oscillate at the higher frequency $\Omega_{\mathrm{R}}$ for $\tau<0$. See \cite{Xu} for a more precise formulation of the above retrodictive argument in terms of the past quantum state formalism \cite{Gammelmark}.

In our calculations, the second atom perturbs the first atom by the Rydberg interaction, and we see that the amplitude of the oscillations of the correlation function is reduced when $V_{12}$ is increased, mainly because of the normalization by the increased mean intensity. Part (b) of Fig. 3, shows the behavior of the cross correlation function between photon counting of the emission by one atom and homodyne detection of the emission by the other atom. In the absence of interactions, there should be no such correlation, and the function plotted should be identical to unity for all times, as shown by the black solid curve in Fig. 3(b). This is clearly not the case for finite $V_{12}$, and from our analysis of the antibunching in Fig. 2  we understand, how detection of a photon from the first atom, leaves the second atom with a finite population in the state $|2\rangle$, which gives rise to the oscillatory evolution of its observables at frequency $\Omega_{\mathrm{R}}$---for both positive and negative $\tau$. The signal amplitude shows a modulation at $\Omega_{\mathrm{R}}/2$, which we ascribe to a finite dark state amplitude of the atom and due to the interaction with the transiently evolving Rydberg state population of Atom 1.

\section{Three-time correlation functions}

In the previous section, the correlations between a field amplitude signal and earlier and later counting signals was qualitatively explained by the conditioned evolution of a monitored quantum system. In this section we elaborate on the formalism of conditional dynamics of a light-emitting system and we review a recently developed theory of past quantum states to interpret our numerical results for three-time intensity correlations and intensity--amplitude--intensity correlations.
\subsection{Conditioned dynamics and the past quantum state}
\label{IIIA}

When a system is observed continuously or at selected instants of time, the density matrix at time $t$, $\rho_{\mathrm{c}}(t)$ is conditioned on the outcome of all measurements performed prior to $t$. The solution to Eq. (1) which describes the \textit{un-conditioned} evolution of a system is therefore modified with the inclusion of stochastic terms to take into account the back action due to the random measurement outcomes. The immediate consequence is that the probability for a given measurement outcome is conditioned on the earlier measurement outcomes, and it is this back action dynamics that manifests itself in the temporal correlation functions in optical detection.

The same correlation also formally makes the probabilities for the early outcomes depend on the later events, and if a system has been monitored on a time interval $[t_1,t_2]$, any measurement at $t \in [t_1,t_2]$ will have outcomes with probabilities that depend on both the prior and posterior measurement results.
The prior information is accounted for by $\rho_{\mathrm{c}}(t)$, while the posterior information is incorporated in an auxiliary matrix, $E(t)$ \cite{Gammelmark}, which equals the identity, $I$,  at time $t_2$, and which solves an equation backward in time, which is the adjoint of the equation for $\rho_{\mathrm{c}}$. For our purpose, it is worth noticing that the adjoint of a quantum jump, $\rho_{\mathrm{c}} \rightarrow \sigma_{12}^i \rho_{\mathrm{c}} \sigma_{21}^i$, is an upward transition transforming $E$ to the excited state just prior to the detection event, $E \rightarrow \sigma_{21}^i E \sigma_{12}^i$, while in intervals with no probing (corresponding to the time between $t$ and $t+\tau$ in our correlation function calculations), $E$ solves an equation like the master equation (1), but backwards in time and with the first of the damping terms in (2) replaced by the Hermitian adjoint $\sum_{j,k} C_k^{(j)\dagger} E  C_k^{(j)}$.

A general measurement is described by the theory of positive operator valued measures, (POVMs), i.e., a set of operators $\{\Omega_m\}$, that fulfils $\sum_m \Omega_m^\dagger \Omega_m = I$. The different operators are associated with different outcomes, enumerated by the continuous or discrete index $m$, and for a given density matrix $\rho(t)$, they yield the outcome probabilities, $P(m)=\mathrm{Tr}(\Omega_m \rho(t) \Omega_m^\dagger)$. With the incorporation of both prior and posterior knowledge, this expression is replaced by \cite{Gammelmark},
\begin{equation} \label{pqs}
P_P(m)=\frac{\mathrm{Tr}(\Omega_m \rho_{\mathrm{c}}(t) \Omega_m^\dagger E(t))}{\sum_{m'} \mathrm{Tr}(\Omega_{m'} \rho_{\mathrm{c}}(t) \Omega_{m'}^\dagger E(t))}.
\end{equation}
Eq. (\ref{pqs}) yields predictions that have been succesfully compared with the outcome of measurements on a microwave cavity field \cite{ENS} and on a superconducting qubit \cite{SCqubit}.

\begin{figure}[tbp]
\centering%
\includegraphics[bb=61 149 518
738,width=1\columnwidth,keepaspectratio]{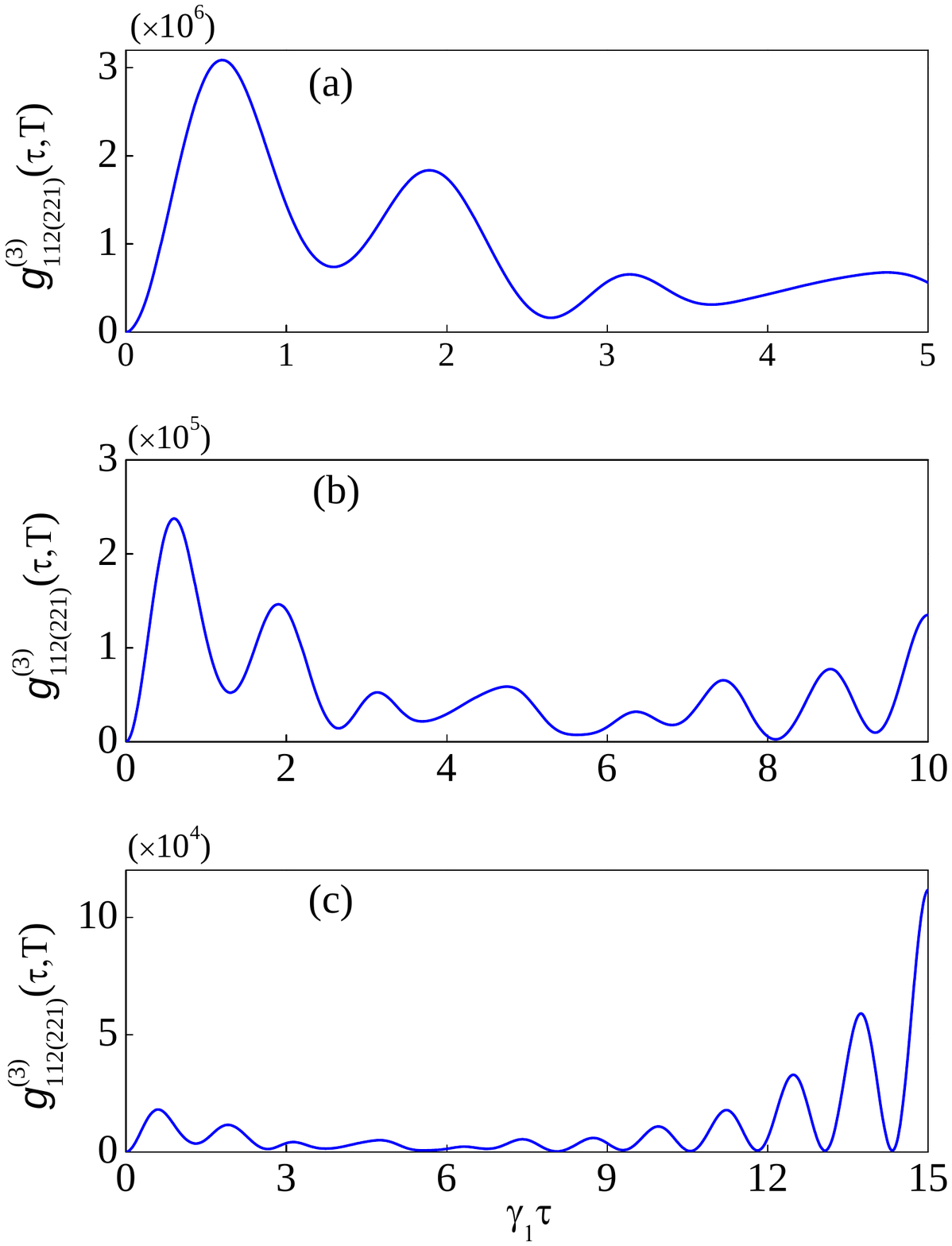}
\caption{(Color online) $g_{112(221)}^{(3)}(\protect\tau ,T)$ on the lower
transitions of two interacting Rydberg atoms for different values of $T=5%
\protect\gamma _{1}^{-1}$ (a), $10\protect\gamma _{1}^{-1}$ (b), and $15%
\protect\gamma _{1}^{-1}$ (c). The other parameters are the same as those in
Fig. \protect\ref{II}.}
\label{gg112}
\end{figure}
\begin{figure}[tbp]
\centering%
\includegraphics[bb=61 149 518
738,width=1\columnwidth,keepaspectratio]{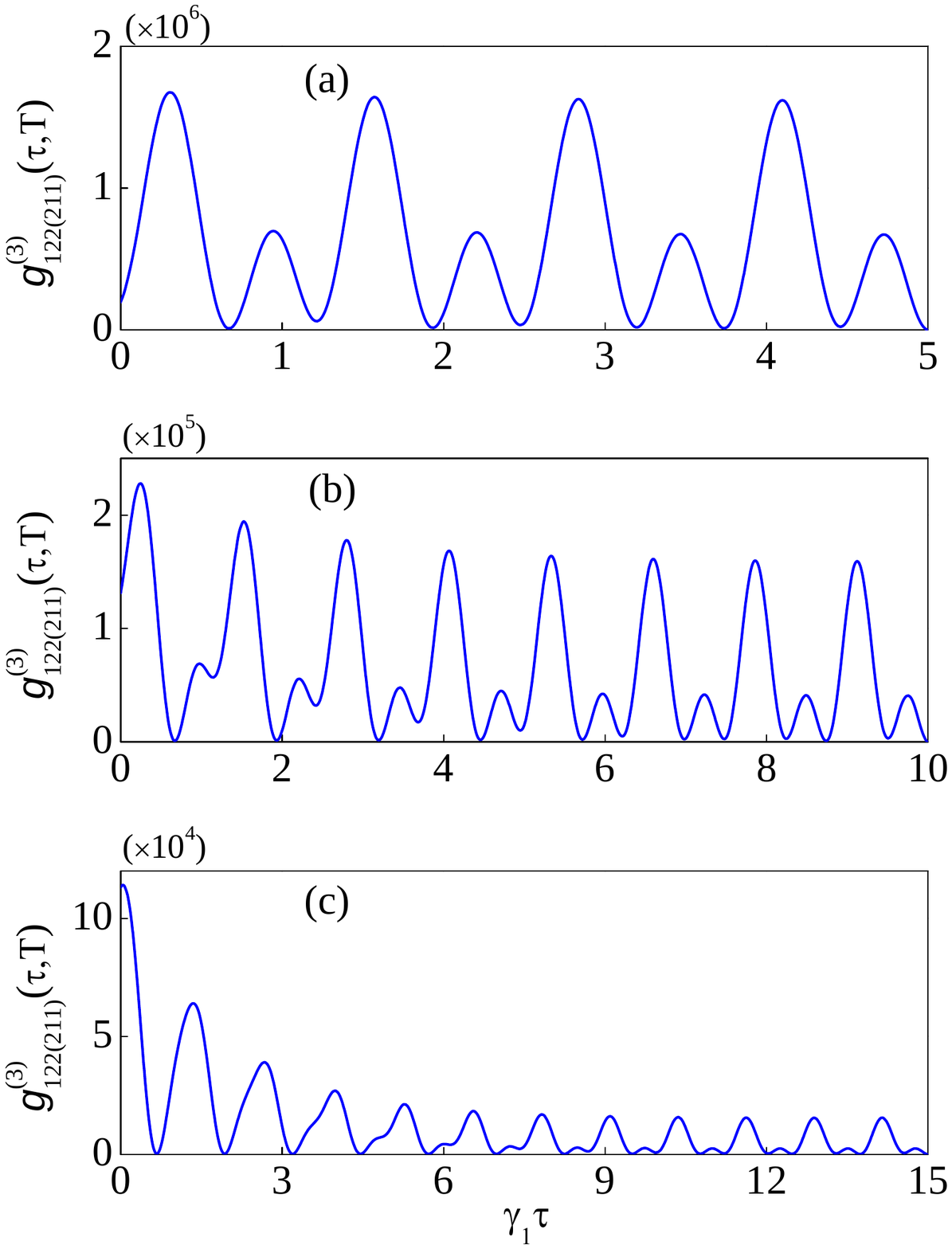}
\caption{(Color online) $g_{122(211)}^{(3)}(\protect\tau ,T)$ on the lower
transitions of two interacting Rydberg atoms for different values of $T=5%
\protect\gamma _{1}^{-1}$ (a), $10\protect\gamma _{1}^{-1}$ (b), and $15%
\protect\gamma _{1}^{-1}$ (c). The other parameters are the same as those in
Fig. \protect\ref{II}.}
\label{gg122}
\end{figure}

\begin{figure}[tbp]
\centering%
\includegraphics[bb=61 149 518
738,width=1\columnwidth,keepaspectratio]{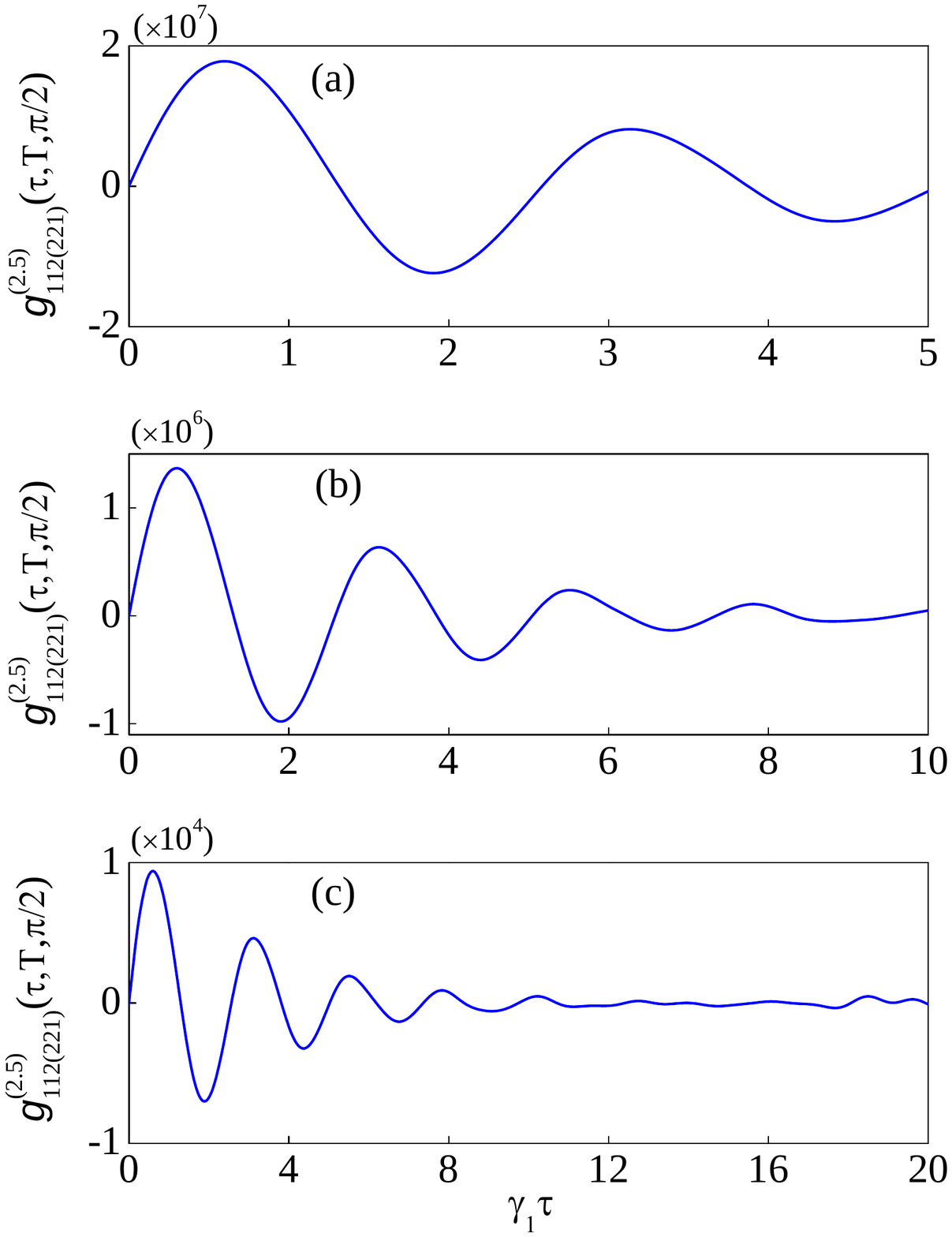}
\caption{(Color online) $g_{112(221)}^{(2.5)}(\protect\tau ,T,\protect\pi %
/2) $ on the lower transitions of two interacting Rydberg atoms for
different values of $T=5\protect\gamma _{1}^{-1}$ (a), $10\protect\gamma %
_{1}^{-1}$ (b), and $20\protect\gamma _{1}^{-1}$ (c). The other parameters
are the same as those in Fig. \protect\ref{II}.}
\label{g112}
\end{figure}
\begin{figure}[tbp]
\centering%
\includegraphics[bb=61 149 518
738,width=1\columnwidth,keepaspectratio]{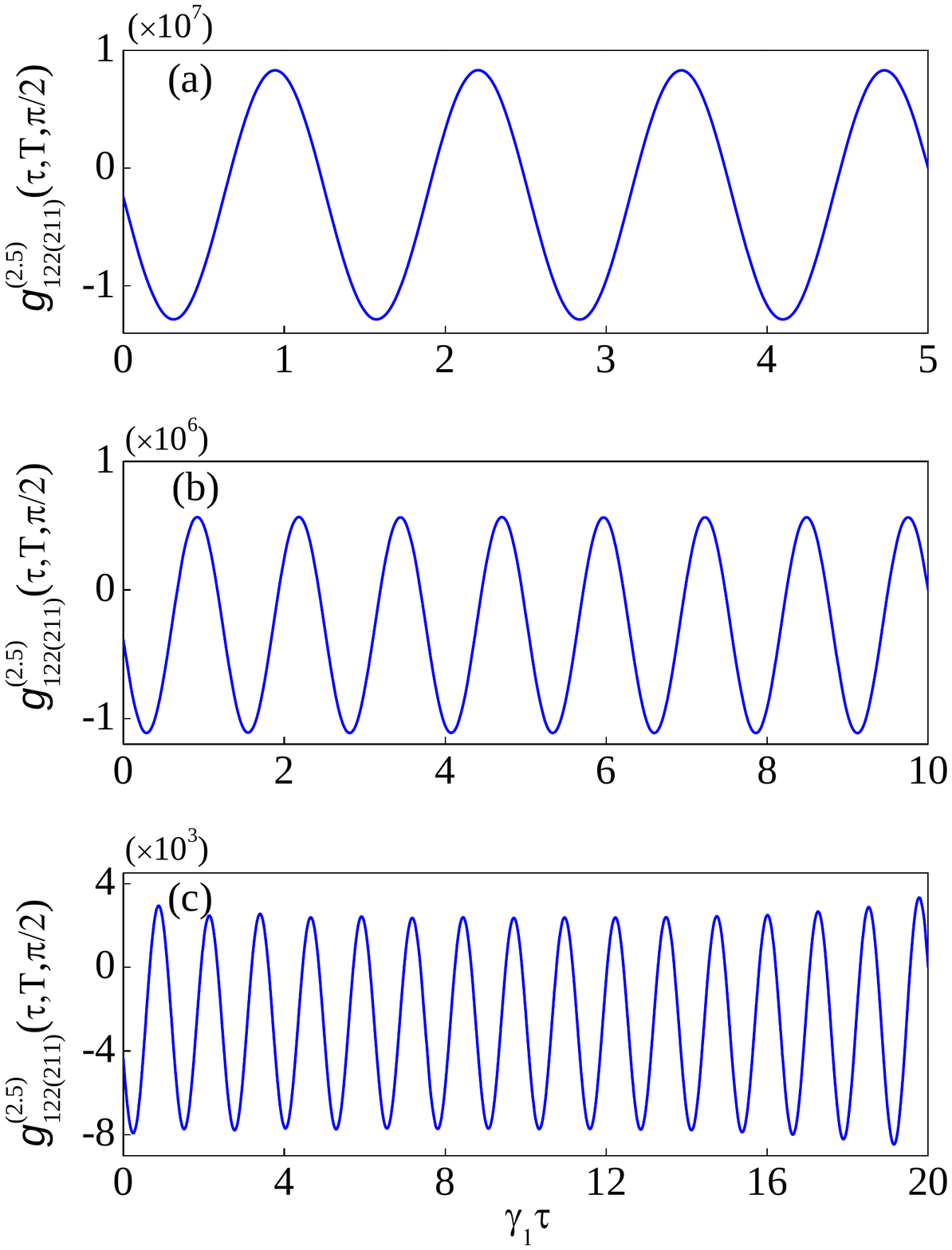}
\caption{(Color online) $g_{122(211)}^{(2.5)}(\protect\tau ,T,\protect\pi %
/2) $ on the lower transitions of two interacting Rydberg atoms for
different values of $T=5\protect\gamma _{1}^{-1}$ (a), $10\protect\gamma %
_{1}^{-1}$ (b), and $20\protect\gamma _{1}^{-1}$ (c). The other parameters
are the same as those in Fig. \protect\ref{II}.}
\label{g122}
\end{figure}
\subsection{Three-time intensity correlation functions}
\label{IIIB}

When applied to intensity and amplitude measurements, Eq. (\ref{pqs}) reproduces the results that we obtained with the quantum regression theorem and illustrated in Figs. (2) and (3), but more importantly, it offers a formal theory from which we can also extract qualitative results and interpretations. We will in this section proceed with an investigation of the three-time correlation function between a prior and a posterior counting event (at times $t$ and $t+T$) and an intermediate intensity or amplitude measurement at time $t+\tau$. We were originally motivated to carry out this study by the apparent conflict between the difference in the evolution frequencies in Fig. 3(a), and the results directed us to a number of interesting observations that we shall present below.

Let us first consider the three-time intensity correlation function,
 \begin{eqnarray}
&&g_{ijk}^{(3)}(\tau ,T)  \notag \\
&=&\lim_{t\rightarrow \infty }\frac{\langle \sigma _{21}^{i}(t)\sigma
_{21}^{j}(t+\tau )\sigma _{22}^{k}(t+T)\sigma _{12}^{j}(t+\tau )\sigma
_{12}^{i}(t)\rangle }{\langle \sigma _{22}^{i}(t)\rangle \langle \sigma
_{22}^{j}(t+\tau )\rangle \,\langle \sigma _{22}^{k}(t+T)\rangle },\qquad
\end{eqnarray}%
with $\tau$ ($T$) being the time interval between the first and the second (third) detector clicks, and $i,j,k=1,2$.

Figure \ref{gg112} shows  $g_{112(221)}^{(3)}(\tau ,T)$ as a function of $\tau$ for different values of $T$. Near $\tau=0$, the function reflects the antibunching of the signal from a single atom, while the ability of two different atoms to emit at the same time is witnessed by the finite value of $g_{112(221)}^{(3)}(T ,T)$.

In comparison, $g_{122(211)}^{(3)}(\tau ,T)$, shown in Fig. \ref{gg122}, reflects bunching for
 $\tau=0$ and antibunching of light coming from the same atom at $\tau=T$. Note that the functions shown in Figs. 4 and 5 are not merely time reversed copies of each other. The oscillatory correlations in Fig. 4 are damped with time, while in Fig. 5, they evolve for long times with only little damping.

\subsection{Three-time intensity--amplitude--intensity correlation functions}
\label{IIIC}

To examine further the evolution of the pair of atoms between the two photon counting events, we supplement the analysis of the three-time intensity correlations with a calculation of the expected outcome of homodyne detection of the field-amplitude between counting events at times $t$ and $t+T$, characterized by%
\begin{eqnarray}
&&g_{ijk}^{(2.5)}(\tau ,T,\theta _{j})  \notag \\
&=&\lim_{t\rightarrow \infty }\frac{\mathrm{Re\,}[\langle \sigma
_{21}^{i}(t)\sigma _{21}^{j}(t+\tau )\,\mathrm{e}^{\mathrm{i}\theta _{j}}\sigma _{22}^{k}(t+T)\sigma
_{12}^{i}(t)\rangle]}{\langle \sigma
_{22}^{i}(t)\rangle \langle \sigma _{22}^{k}(t+T)\rangle \,\mathrm{Re\,}%
[\langle \sigma _{21}^{j}(t+\tau )\rangle \,\mathrm{e}^{\mathrm{i}\theta
_{j}}]},\qquad
\end{eqnarray}%
with $i,j,k=1,2$.

We focus on the same cases of prior and posterior counting events from different atoms while probing now the field amplitude from the atom that caused either the first or the last count. We find that the correlations between the initial intensity measurement on one atom and the subsequent amplitude measurement on the same atom decay after a few oscillation periods in Fig. 6, while the correlations involving the amplitude measurement and the last counting measurement on the same atom are undamped as a function of $\tau$ in Fig. 7.

The persistent oscillations in $g_{122(211)}^{(2.5)}(\tau ,T,\pi/2)$ over time intervals much longer than the intermediate atomic state lifetime $1/\gamma_1$, indeed have a natural explanation in terms of the past quantum state formalism in Eq. (\ref{pqs}). The time evolution of $\rho_{\mathrm{c}}$ between the detector click at $t$ and the amplitude detection at time $t+\tau$ is governed by the usual master equation, with the initial state given by the application of the jump (down) operator on the steady state density matrix. Likewise the matrix $E$ is given by the adjoint master equation solution, propagated backwards in time from the value, post-selected by the counting event at time $T$. Just prior to this counting event, $E$ is given by a jump (up) operator applied to the identity matrix \cite{Gammelmark}.

\begin{figure}[tbp]
\centering%
\includegraphics[bb=76 321 491
678,width=0.95\columnwidth,keepaspectratio]{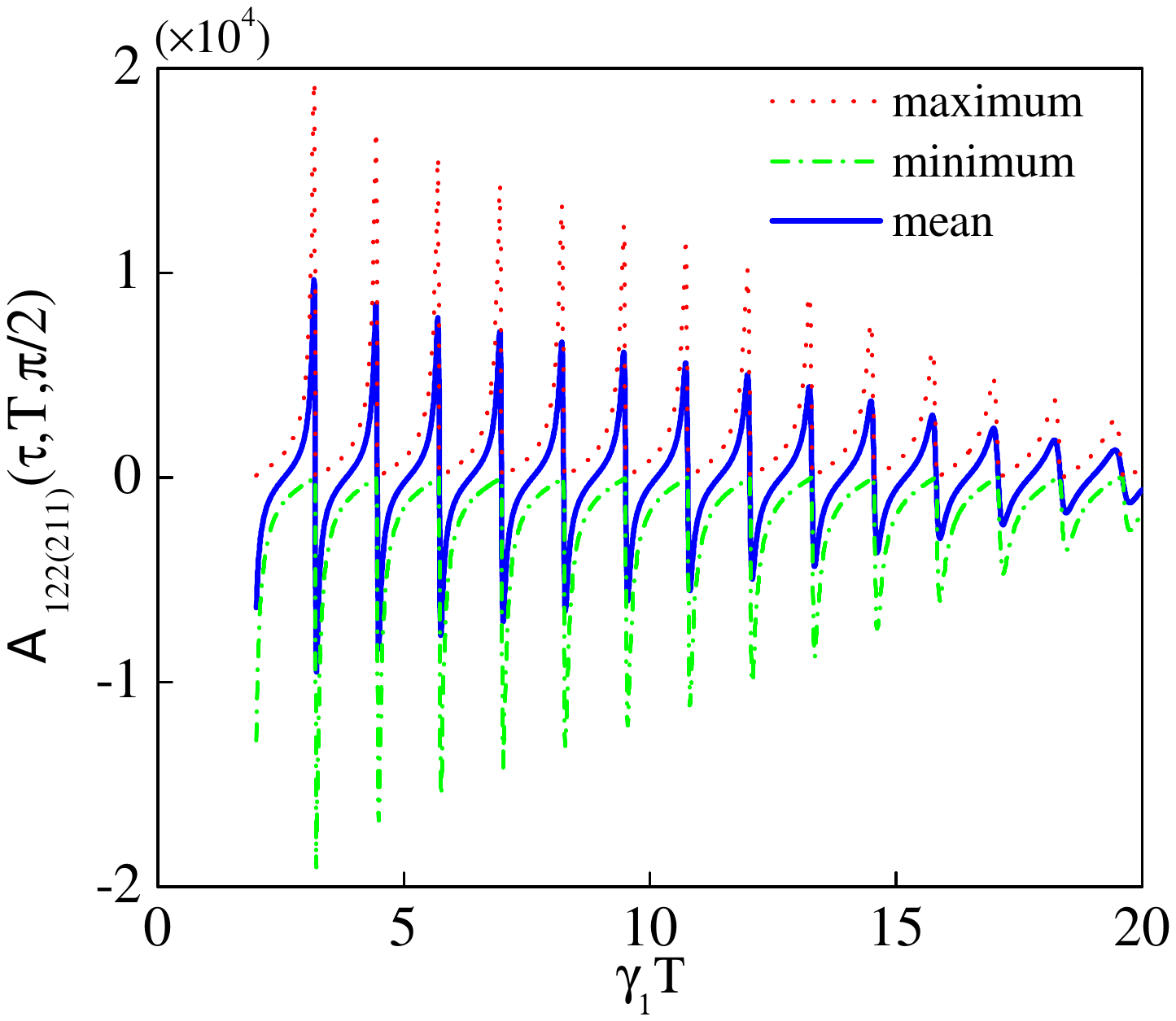}
\caption{(Color online) The maximum (red dotted line), minimum (green
dash--dotted line), and mean (blue solid line) of $\mathcal{A}_{122(211)}(\tau,T,\pi/2)$ around $\tau=T/2$ as a function of $T$. The other parameters are the same as those in
Fig. \protect\ref{II}.}
\label{mean}
\end{figure}

Despite $\rho_{\mathrm{c}}(t)$ and $E(t)$ having quite different dynamics, the linear set of equations for their matrix elements have identical spectra of complex eigenvalues $\lambda_j$ and their time evolution can be expanded on eigensolutions which evolve with complex exponential factors $\exp(\lambda_m \tau)$ and $\exp(\lambda_n(T-\tau))$, respectively. The predictions by
Eq. (\ref{pqs}) will hence be a combination of products of such factors, yielding $\tau$-independent terms (when $m=n$), and undamped, oscillatory $\tau$-dependent terms from eigenvalue pairs with $\lambda_n = x_n+\mathrm{i }y_n= \lambda_m^*$. All complex eigenvalues occur in such pairs since the density matrix equations can be represented by real sets of equations.

The sum of constant and oscillatory terms causes the oscillations in Fig. 7 to occur around constant levels that depend on $T$. We have illustrated this dependence in Fig. 8, where the lower (green dash--dotted) and upper (red dotted) curves show the range of variation of the field amplitude $\mathcal{A}_{122(211)}(\tau,T,\pi/2)$ around $\tau=T/2$ as a function of $T$, normalized by the intensity--intensity correlation between $t$ and $t+T$,
\begin{equation}
\mathcal{A}_{ijk}(\tau,T,\theta_{j})=\frac{g_{ijk}^{(2.5)}(\tau,T,\theta_{j})}{%
g_{ik}^{(2)}(T)}.
\end{equation}%

We observe that even for very long intervals $T$ between the clicks, for one value of $T$, the amplitude oscillates between zero and a large positive value, while for an only slightly larger $T$, the oscillations occur between a correspondingly large negative value and zero. These abrupt changes follow the periodicity of the intensity--intensity correlation function, shown in Fig. 2, which implies that they are rare and difficult to observe.


While the persistent oscillations in Figs. 5 and 7 follow from the above eigenvalue argument, we need to explain why the oscillations in Figs. 4 and 6 are damped. The results in Figs. 4--7 are all obtained by solution of the same differential equations, but the conditioning counting events impose boundary conditions that strongly influence the observed dynamics. After counting a photon from Atom 1, the density matrix factorizes into a product state $ \rho \rightarrow \sigma_{12}^{(1)} \rho \sigma_{21}^{(1)} \propto |1\rangle_1 \langle 1|\otimes \rho_{\mathrm{c},2}$, where the conditioned state $\rho_{\mathrm{c},2}$ of Atom 2 depends on the steady state of the joint system, attained prior to the jump. Just before the counting of a photon from Atom 2 at time $t+T$, the matrix $E$ acquires the form $ E \rightarrow \sigma_{21}^{(2)} I \sigma_{12}^{(2)} \propto  I_1 \otimes |2\rangle_2 \langle 2|$, \textit{i.e.,} it factorizes into a product of the identity matrix $I_1$ acting on the Atom 1 Hilbert space and the excited state projector of Atom 2.

To understand why these limiting conditions lead to the observed correlations, we consider for simplicity the evolution between $t$ and $t+T$ in the absence of the Rydberg interaction. The matrices pertaining to the separate atoms then evolve independently, and while the three terms, $|1\rangle_1 \langle 1|$, $\rho_{\mathrm{c},2}$ and $|2\rangle_2 \langle 2|$ all develop damped oscillatory dynamics, the identity matrix $I_1$ is invariant under the backward time evolution of the matrix $E$. This, in turn, implies that according to Eq. (\ref{pqs}) the outcome probabilities for measurements on Atom 1 between 0 and $T$ are given by the usual density matrix expression and hence they undergo the usual damping towards steady state. In our argument about eigenvalues, the terms in the evolution of $\rho$, damped as $\exp(\lambda_m \tau)$, are not multiplied by the matching functions $\exp(\lambda_n (T-\tau))$ because $I_1$ is the zero eigenvalue solution of the backward evolution equation. The interaction between the atoms of course modifies the argument, but the solutions change and the factorization becomes invalid only gradually as the equations are integrated over time.

\section{Discussion}

In summary, we have studied photon correlations in signals emitted by a pair of interacting atoms. The interaction significantly alters the dark states of the individual atoms, and the optical response is correspondingly large. This may pave the way for the use of photon correlation measurements as a probe of the Rydberg interaction strength, and, ultimately we may use the detection of fluorescence from one atom to herald the emission of a nonclassical state of light by the other atom.

The correlations reveal a host of interesting results. In particular, we find strong correlations between counting events on one atom and field amplitude measurements on the same and on the other atom. We observe a strong asymmetry between negative and positive time correlations and ascribe this to the different boundary conditions for the equations of motion: An earlier photon count is accompanied by a jump into the ground state, while a later count event specifies the matrix $E$ to populate the short-lived excited state, and these states have different expansions on the eigenstates of the ladder system Hamiltonian.

Three-time correlation functions show oscillatory behaviour as function of the intermediate time argument $t+\tau$ between counting events at times $t$ and $t+T$, and these turn out to be undamped for times much longer than the lifetimes of the atomic states. This result is a consequence of the  post selection by the last detection event, and it is explained quantitatively by the contribution of exponentially damped terms with both time arguments $\tau$ and $T-\tau$. Interestingly, due to the conditions imposed by the detection events, this phenomenon occurs for some but not all combinations of correlation functions of the emission signals from two atoms.

Our work focused on pairs of atoms, but we imagine that coherent ensemble emission from Rydberg blocked ensembles and from atoms with more complex level structure (e.g., excited from a larger Zeeman and hyperfine ground state manifold) \cite{Line}, will show similar behavior and may lead to applications of the results presented.

\bigskip \textbf{Acknowledgements}

The authors acknowledge useful discussions with David Petrosyan, Eliska Greplova, Etienne Brion, and Brian Julsgaard, and
financial support from the Villum Foundation and the IARPA MQCO program.

\end{document}